\documentclass[fleqn,usenatbib]{mnras}
\usepackage[T1]{fontenc}
\usepackage{ae,aecompl}
\usepackage{graphicx}
\usepackage{soul}

\usepackage{amssymb}
\usepackage{amsmath}
\usepackage{float}
\usepackage{times}
\usepackage[markup=default]{changes}
\usepackage{soul}

\usepackage{color}

\newcommand{\kms}{\,km\,s$^{-1}$}

\title[Structure of  accretion flows in  nova-like CVs]{The structure of  accretion flows in  nova-like cataclysmic variables: \\  RW Sextantis and 1RXS J064434.5+334451 }

\author[M. S. Hernandez et al.]{
M. S. Hernandez$^{1,2}$,  S.  Zharikov$^1$,  V. Neustroev$^{3,4}$ and G. Tovmassian$^1$
\\
$^{1}$Instituto de Astronom{\'i}a, Universidad Nacional Aut{\'o}noma de M{\'e}xico, Apdo. Postal 877,  Ensenada,  Baja California, M{\'e}xico , 22800    \\
$^{2}$ Instituto de F\'{i}sica y Astronom\'{i}a, Facultad de Ciencias, Universidad de Valpara\'{i}so, Av. Gran Breta\~{n}a 1111 Valpara\'{i}so, Chile \\
$^{3}$Finnish Centre for Astronomy with ESO (FINCA), University of Turku, V\"{a}is\"{a}l\"{a}ntie 20, FIN-21500 Piikki\"{o}, Finland \\
$^{4}$Astronomy research unit, PO Box 3000, FIN-90014 University of Oulu, Finland}

\begin{document}

\date{Accepted 2017 May 26. Received 2017 May 17; in original form 2017 March 22}
\pubyear{2017}

\pagerange{\pageref{firstpage}--\pageref{lastpage}} \pubyear{2015}

\maketitle

\label{firstpage}

\begin{abstract}
New time-resolved optical spectroscopic echelle observations of the
nova-like cataclysmic variable RW~Sextantis  were obtained, with the aim to
study the properties of emission features in the system.  The profile of
the H$\alpha$ emission line can be clearly divided into two (`narrow' and
`wide') components. 
Similar emission profiles are observed in another nova-like system, 1RXS~J064434.5+33445,  for which  we also  
reanalysed the spectral data and redetermined the system parameters. 
The source of the `narrow', low-velocity component is the irradiated face 
of the secondary star.   We disentangled and  removed the `narrow' component from the H$\alpha$
profile to study the origin and structure of the region emitting the
wide component.   
We found  that the `wide' component  is not related to the white
dwarf or the wind from the central part of the accretion disc, but is emanated from the outer side of the disc.
Inspection of literature on similar systems 
 indicates 
 that this feature is common for some other long-period nova-like variables.
We propose that the source of the `wide' component is an extended, low-velocity region in the outskirts 
of the opposite side of the accretion disc,  
with respect to the collision point of the accretion stream and the disc.

\end{abstract}



\maketitle

\section{Introduction}
\label{sec:intro}
Cataclysmic variables (CVs) are interacting binaries comprised
of a white dwarf (WD) as the primary and a late-type (K--M type) main-
sequence star or a brown dwarf as the secondary \citep{Warner:1995aa}. The
secondary star fills  its Roche lobe and loses matter via the inner Lagrangian
point L$_1$. Infalling matter forms an accretion disc around the WD
unless its magnetic field is strong enough to prevent this. In CVs with 
low mass-transfer rates $\dot{M} < 10^{-9} M_{\sun}$, known as dwarf novae (DNe), the accretion disc is 
usually in the low temperature and surface density  state (quiescence).  
Every once in a while, the accretion disc 
builds up by increasing both the  temperature and the surface density reaching  the thermal instability, which results in regular outbursts.
Nova-like variables (NLs)  in contrast to DNe are CVs that do
not display such  eruptive activity.  
The mass-transfer rate in NLs is estimated to be
$\dot{M}\gtrsim 10^{-9} M_{\sun}$ \citep{Warner:1995aa}
 which allows  their discs
to remain in a high state most of the time. Spectrally, they resemble DNe in outbursts. They are found to have orbital periods
longer than $\gtrsim 2$ hours.   
Historically, NLs that do not exhibit magnetic characteristics are also called  UX~UMa-type stars.

The optical spectra of NLs have a blue continuum and show a wide range of appearances,
from pure  absorption-line spectra to pure  emission-line spectra. In some objects, the emission
lines are embedded in broad absorption troughs. A typical NL spectrum usually shows a set of Balmer,
\ion{He}{I} and \ion{He}{II} lines, and also the Bowen blend of fluorescence lines and weak absorption lines of metals.
The profiles of emission lines in the optical range are usually
single-peaked regardless of the system's inclination.
This is in contrast  to DNe, where emission lines are practically always double-peaked in high-inclination
systems. The UV spectra of NLs exhibit P Cygni profiles of emission lines, implying the presence of a strong high-velocity
wind ($\sim$3000 \kms, see, e.g. \citealt{Noebauer:2010aa}) in the
inner part of the accretion disc. 

Many NLs show long-term brightness variations on different time-scales. The
most outstanding behaviour is observed in VY~Scl stars (or `anti-dwarf novae'), whose
light curves display occasional, unpredictable low states of $\sim$1--6 mag for several weeks
or even years. There is also another large group of systems, called SW Sex stars \citep{Thorstensen:1991aa},
which is distinguished from the rest of NLs by common properties.
The SW~Sex stars are mostly concentrated in the $\sim$3--4~h orbital period range.
Most of them are deeply eclipsing systems, showing
odd V-shaped eclipse profiles and simple single-peaked emission lines. They also show a distinct phase
shift between the eclipse and the zero phase of the radial-velocity curve, high-velocity emission S-waves
with maximum blueshift near phase $\sim$0.5 and transient absorption dips in the emission lines around
phase 0.4--0.7.

In order to explain the appearance of single-peaked emission-line profiles in high-inclination CVs,
a number of mechanisms have been widely discussed in the literature \citep[see e.g.][]{Dhillon:1997aa}.
In this paper, we make a new attempt to probe the formation of such lines,
based on high-resolution spectroscopy of two  long-period NLs,
RW~Sextantis and 1RXS~J064434.5+334451, for which hints of a multicomponent structure of emission lines in
low-resolution spectra have been reported.

\begin{figure*}
\setlength{\unitlength}{1mm}
\resizebox{11cm}{!}{
\begin{picture}(100,105)(0,0)
\put (-20,0)  { \includegraphics[width=14.cm,  bb = 20 40 705 559,clip=, ]{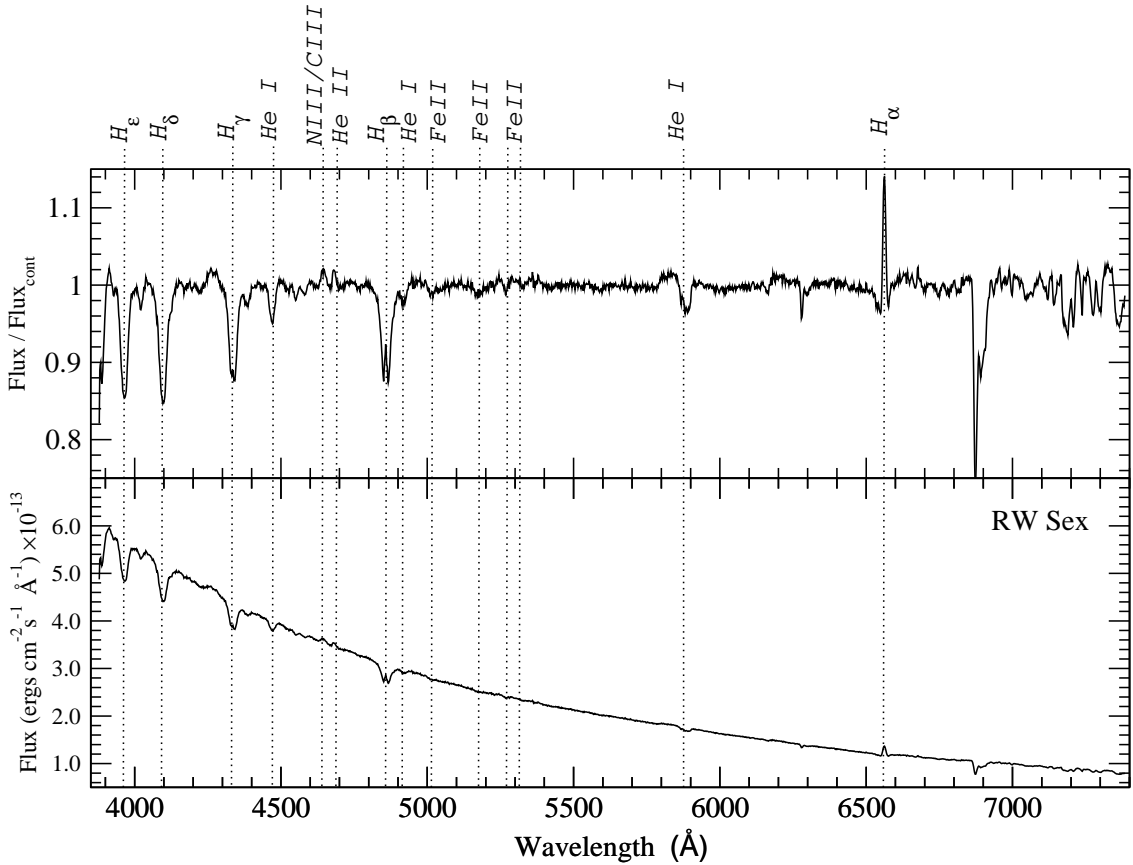}}
\end{picture}}
  \caption{The flux-calibrated (bottom) and continuum-normalized (top) medium-resolution 
           spectra of RW Sex. The spectral features are marked.}
  \label{fig1:spec}
\end{figure*}

RW~Sex is a relatively bright CV (V$\sim$10.6 mag) with an orbital period of
$P_{\tt orb}$ = 0.24507 d \citep{1992A&A...256..433B}.
The  optical spectrum of RW~Sex shows broad  Balmer absorption lines with
multicomponent emission cores, and weak \ion{He}{II} and \ion{C}{III}/\ion{N}{III} emission
lines (see Fig.~\ref{fig1:spec}).  Based on phase-resolved spectroscopy,
\cite{1992A&A...256..433B} provided estimates of the
RW~Sex component  masses and the orbital inclination: $M_1 = 0.84M_{\sun}$,
$M_2=0.62M_{\sun}$, $i = 34\degr\pm6$, and the mass ratio $q\equiv M_2/M_1 = 0.74$. 
They also estimated the distance to the system of 150~pc. This is inconsistent
with the Hipparcos parallax of 3.46$\pm2.44$ mas \citep{Perryman:1997aa},
which corresponds to a distance of $289^{+689}_{-119}$~pc.
The mass-transfer rate is estimated at $\dot{ M} \,= \,(0.3 - 1.0) \times \, 10^{-8} \, M_{\sun}$\, yr$^{-1}$
\citep{2010ApJ...719..271L,Vitello:1993aa, 1982ApJ...258..209G}. Ultraviolet
observations show P~Cyg profiles of emission lines \citep{1982ApJ...258..209G, 2003MNRAS.340..551P},
indicating the presence of a hot, fast (up to 4500~\kms) wind from the innermost portions of the disc.
\citet{2010ApJ...719..271L} analysed \textit{Far Ultraviolet Spectroscopic Explorer, Hubble Space Telescope} and \textit{International Ultraviolet Explorer} spectra of
RW~Sex and found the following system parameters:  $M_1=0.9 M_{\sun}$,
$M_2=0.67 M_{\sun}$, $\dot{M}=2.0\times 10^{-9} M_{\sun}$\,yr$^{-1}$, $i =
34\degr$,  $T_{\tt WD}$=50\,000K.  \citet{Coppejans:2015aa} reported the Very Large Array radio detection of RW~Sex.  The object showed  non-variable flux  of
$\sim$33.6~$\mu$Jy beam$^{-1}$ in the range  of 4226--8096 MHz with  a spectral  index of $\alpha = -0.5\pm0.7$ ($F = \nu^\alpha$), and probably of non-thermal origin.
A compilation of available infrared  data from the Two-Micron All-Sky Survey, {\it Spitzer} and the {\it Wide-field Infrared Survey Explorer} was reported by \citet{Hoard:2014aa} who  discussed the possible origin of the infrared excess in terms of emission from bremsstrahlung or circumbinary dust, with either mechanism facilitated by the mass outflows (e.g., disk wind/corona, accretion stream overflow, etc.) presented in NLs.

 \begin{figure}
\setlength{\unitlength}{1mm}
\resizebox{11cm}{!}{
\begin{picture}(100,50)(0,0)
\put (0,0)  {  \includegraphics[width=7.25cm,  bb = 40 100 705 525,clip=]{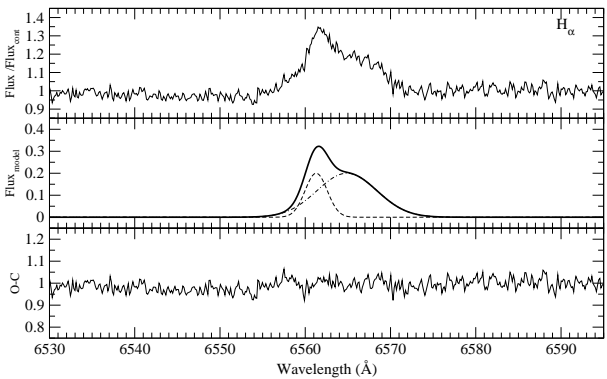}}
\end{picture}}
  \caption{An example of the high-resolution H$\alpha$ profile of RW~Sex (top panel),
           the result of a double-Gaussian fit to the profile  (middle panel)
           and the residuals between the observed and 
           calculated profiles (bottom panel).
The  underlying broad absorption component was removed by continuum normalisation.}
  \label{fig:Ha}
\end{figure}

1RXS J064434.5+334451 ($V$$\sim$13.3 mag) was discovered by \citet{Wozniak:2004aa} in the
Northern Sky Variability Survey.
\citet{Sing:2007aa} reported this object as a deep eclipsing CV with
the orbital period of  
$P_{\tt orb}$ = 0.26937d
and  derived the physical parameters of the
system, such as the WD mass $M_1= 0.66 M_{\sun}$, 
the mass ratio $q=0.78$
and the WD temperature $T_{\tt WD}$$\sim$25\,000~K.
They classified the system as a NL CV of the UX~UMa-type or the SW~Sex-type.
\citet{Hernandez-Santisteban:2012aa}, \citet{Echevarria:2015aa}, and \citet{Hernandez-Santisteban:2017aa} recently
reported new time-resolved photometry and echelle spectroscopy of 1RXS~J064434.5+334451.
They constructed Doppler maps of the system in the H$\alpha$, H$\beta$,
and \ion{He}{II} 4686 emission lines and refined system parameters based
on the obtained data. The WD with $M_1=0.82\pm0.06M_{\sun}$ and the mass ratio $q=0.96\pm0.05$  were proposed assuming the system inclination of $i$=78$\degr\pm2$ in \citet{Hernandez-Santisteban:2017aa}  that is in  
a disagreement with  \citet{Sing:2007aa}  and their own recent estimations: $M_1 = 0.91\pm0.04M_{\sun}$, $q=0.91\pm0.09$ from \citet{Hernandez-Santisteban:2012aa} and 
$M_1 = 0.76\pm0.04M_{\sun}$, $q=0.75\pm0.1$ from \citet{Echevarria:2015aa}.

Regardless of different inclination angles, the Doppler maps of  Balmer lines of  1RXS~J064434.5+334451 and RW~Sex
are very similar. The H$\alpha$ profile  appears to have multiple
but identical  components in both objects.  Therefore,  we analyzed  the profiles of intense emission lines
of both objects side by side in order to understand the origin and the sources of their constituent
components. Moreover, we also redetermined  the system parameters of  1RXS~J064434.5+334451 based on the available time-resolved photometry and  eclipse profile fitting.
In this study, we used  new observations of RW~Sex and the original  data of 1RXS~J064434.5+334451 obtained and kindly provided
by  Hern\'{a}ndez~Santisteban et al.

This paper is structured as follows. In  Section~\ref{sec:obs}, we
describe our  spectroscopic observations of RW~Sex and the
corresponding data reduction.  In Section~\ref{RWSexDopmap}, we present
an analysis   of emission-line profile components  based on Doppler tomography of  RW Sex.
In Section \ref{RXSDopmap}, we revisit 1RXS~J064434.5+334451
to determine the system parameters 
and to reanalyse its Doppler maps.
The general discussion of obtained results and their application to NL
systems of similar orbital periods follows in Section~5. Our conclusions
are presented in Section~\ref{conclud}.

\begin{figure*}
\setlength{\unitlength}{1mm}
\resizebox{11cm}{!}{
\begin{picture}(100,155)(0,0)
\put (-20,0)  {  \includegraphics[width=14cm,  bb = 20 150 550 730,clip=]{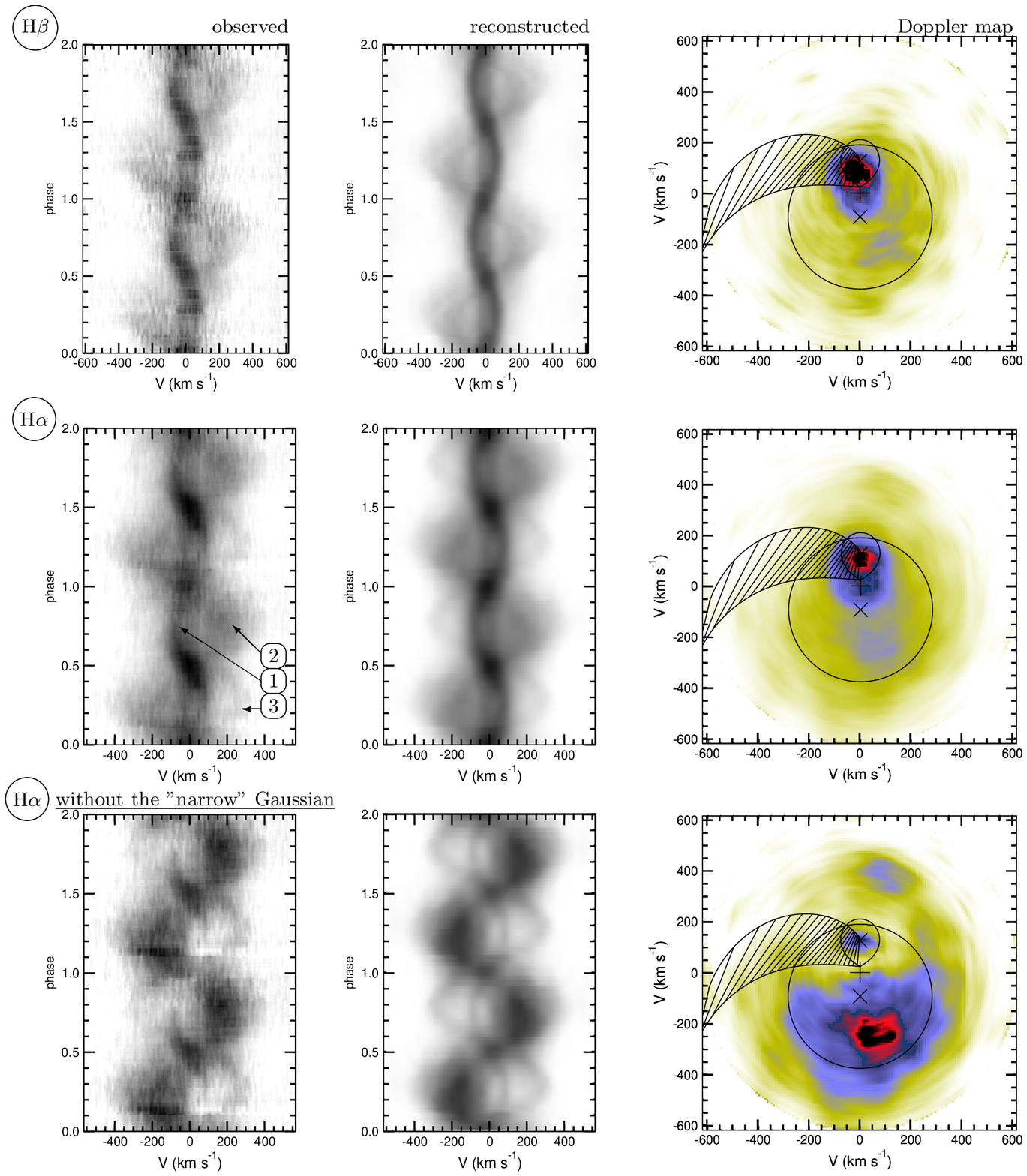}}
\end{picture}}
  \caption{The observed and reconstructed trailed spectra and the corresponding Doppler maps
           (from the left to right columns) of the H$\alpha$ and H$\beta$ emission lines of RW~Sex.
           The two upper rows show unaltered
           observed lines, while the bottom row shows the H$\alpha$ line after the
           subtraction of the `narrow' emission component. Indicated on the maps are the positions of the
           Roche lobe of the secondary (upper bubble with the cross), the centre of mass of the binary
           (middle plus symbol) and the WD (lower cross). The trajectory of the gas stream and the Keplerian
           velocity of the disc along the stream are also shown in the form of the lower and upper
           curves, respectively. The circle in the Doppler maps shows the tidal limitation radius {\tt r$_d$(max)}
           of the accretion disc. All the marks are plotted for $M_1=0.84$ M$_{\sun}$,
           $i=34\degr$ and $q=0.74$. The `narrow', wide and third component of emission lines are denoted
           in one of the panels by respective numbers.
   }
  \label{fig:dopHa}
\end{figure*}
  \begin{table}
 \centering
    \caption{Log of RW Sex spectroscopic observations. Echelle denotes the REOSC echelle spectrograph and B\&Ch denotes the long-slit Boller and Chivens spectrograph.}

\begin{tabular}{ccccc} \hline
Date      &   HJD start& Number  &  Exposure &  Duration       \\
DD/MM/YYYY & +2450000 & of exp. & (s) & (h) \\ \hline
Echelle        &   &               &                        &                   \\
13/01/2015 &     7036.03109  &   3      &      900-1200  &           0.9 \\
14/01/2015 &     7036.83745  &   10     &       900             &               2.5\\
15/01/2015&      7037.83663  &  10      &       900-1200        &               2.8\\
16/01/2015&      7039.00118   & 6       &       900             &               1.5\\
17/01/2015&      7039.79971  &  15      &       900             &               3.7\\
18/01/2015&      7040.79247  &  15      &       900             &               3.7\\
19/01/2015&      7041.79253 &   14      &       900             &               3.5\\
23/03/2016&     7470.83605 &    3       &       1200            &               1.0\\
24/03/2016&     7471.78651 &    6       &       1200            &               2.0\\
25/03/2016&     7472.77510&     7       &       1200            &               2.3\\
26/03/2016&     7473.80031&     6       &       1200            &               2.0\\
27/03/2016&     7474.77762&     6       &       1200            &               2.0\\
27/03/2016&     7475.78476&     8       &       1200            &               2.7\\  \hline
B\&Ch        &&&& \\
19/01/2016& 7406.85160  &       1       &       1200            &               0.33\\
18/01/2016&7407.87406   &       2       &       1200            &               0.66\\
10/02/2016& 7428.79462  &       20      &       300             &               1.65\\
30/03/2016& 7477.64920 &      31      &     900            &             5.7  \\

\hline

\end{tabular}

\label{tab:log}
\end{table}

\begin{figure*}
\setlength{\unitlength}{1mm}
\resizebox{11cm}{!}{
\begin{picture}(100,142)(0,0)
\put (-20,0)  {  \includegraphics[width=13cm,  bb = 90 290 490 720,clip=]{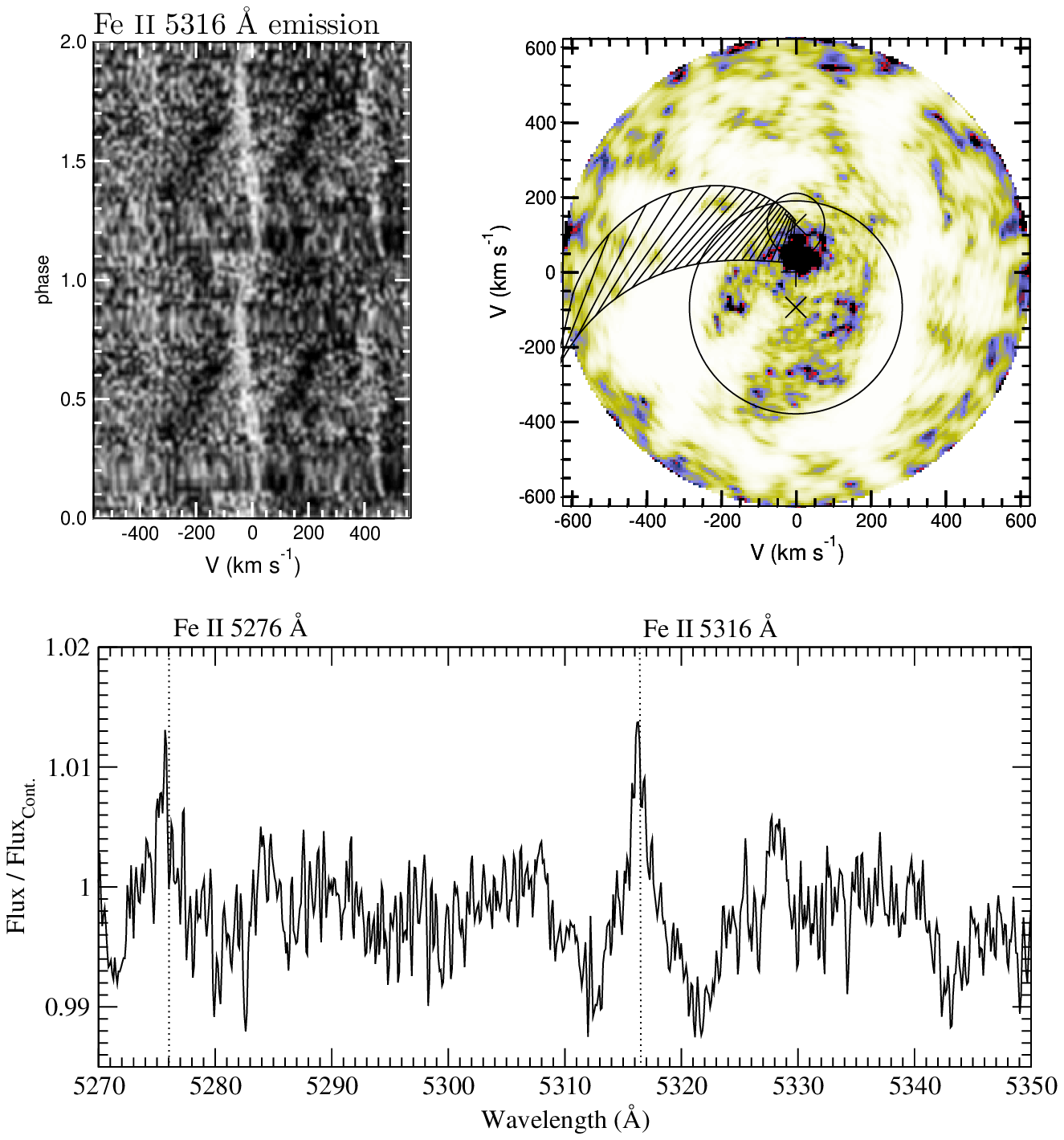}}
\end{picture}}
  \caption{
   Bottom: the averaged spectrum of RW~Sex in the 5270--5350 \AA\ wavelength range, uncorrected
   for orbital motion.
    Top: the trailed spectrum of the \ion{Fe}{II} 5316 \AA\ line  (left) and the corresponding
    Doppler map (right). The symbols on the Doppler map are overplotted in the same way as for
    Fig.~\ref{fig:dopHa}.}
 \label{fig:FeII}
\end{figure*}

\section{Observations and Data Reduction}
\label{sec:obs}

The spectroscopic data of RW~Sex were obtained using the echelle REOSC
spectrograph \citep{Levine1995}    attached to  the 2.1-m telescope of
the Observatorio Astron\'omico Nacional at San Pedro M\'artir (OAN
SPM)\footnote{http://www.astrossp.unam.mx}, Mexico.  The echelle
spectrograph provides spectra spread over 27 orders, covering the
spectral range $\sim$3500--7105~\AA\ with the spectral resolving power of
R$\approx$18000. A total of 73 echelle spectra were obtained during
seven consecutive nights in 2015 and 36 spectra in 2016. A Tr-Ar lamp was
used for wavelength calibration. The spectra were reduced using the {\it
echelle} package in {\sc iraf}\footnote{IRAF is distributed by the National Optical Astronomy Observatories, which
are operated by the Association of Universities for Research in Astronomy, Inc., under cooperative agreement with the National Science Foundation.}. The medium resolution ($\sim$1.15~\AA~pix$^{-1}$)
spectra were obtained using the same 2.1-m telescope and the Boller and
Chivens  (B\&Ch) spectrograph. Standard procedures, including bias and flat-field
correction, cosmic ray removal, wavelength and flux calibration were applied
using  the corresponding tasks in {\sc iraf}.
The log of observations is given in  Table~\ref{tab:log}.

Spectra of 1RXS J064434.5+334451  were obtained by  \citet{Hernandez-Santisteban:2012aa} and 
\citet{ Hernandez-Santisteban:2017aa} 
using the same  instrument/telescope set-up as in the case of RW Sex, at the same observatory. Moreover, they
 obtained  the time-resolved photometry of the object with the 1.5-m telescope  also located at OAN SPM.
The full description of those observations and data reduction has been presented in previously  cited publications. 
Only the H$\alpha$ order of echelle spectra was used in our following analyses.

\section{The spectrum, Balmer emission lines and Doppler tomography of RW Sex}
\label{RWSexDopmap}

Fig.~\ref{fig1:spec}  shows the medium-resolution spectrum of RW~Sex. The
spectrum has strong broad Balmer absorption lines with multicomponent
emission cores that are clearly visible in H$\alpha$ and partly visible in H$\beta$, and there is a
hint of emission in H$\gamma$. Higher Balmer series lines show only
absorption profiles. There are also \ion{He}{I} 4026, 4144, 4388, 4471, 4922,
5015, 5876 and very weak  \ion{Fe}{II} 5169, 5276,  5316 absorption/emission lines.
The high-excitation lines of \ion{He}{II} 4686 and the \ion{C}{III}/\ion{N}{III} 4634--4651
blend are detected in emission. Also, the spectrum shows a relatively strong emission
bump centred at $\sim$5830 \AA, which is probably a blend of \ion{C}{III} 5827 and the
\ion{C}{IV} lines at 5801 and 5812 \AA, and possibly of other highly excited lines
blueward
of \ion{He}{I} 5876. These lines are very uncommon for CV spectra,
their detection in this wavelength region has been reported only for a few CVs
\citep[see, e.g.,][]{NeustroevSSS}.
The continuum in a wide optical range can be described by a power law
$F_\lambda \sim \lambda ^{\alpha}$, where $\alpha = - 3.08\pm0.01$ is steeper than  $\alpha_{\tt st} =-2.33$ adopted for the standard
disc model \citep{Lynden-Bell:1969aa}.

Fig.~\ref{fig:Ha} shows an example of a high-resolution (R$\approx$18000) profile
of the emission core of H$\alpha$. The profile has a complex structure consisting of
at least two distinct variable components, which are clearly visible in trailed spectra
of H$\alpha$ and H$\beta$ (Fig. ~\ref{fig:dopHa}, left-hand panels).
One of these components looks strong  and narrow, and it exhibits low-amplitude radial-velocity variations ($\lesssim$100 \kms), while the other seems weaker and wider and shows
significantly larger radial-velocity variability ($\approx$200--250 \kms).

\begin{table}
 \centering
    \caption{Parameters of the Gaussian components of the H$\alpha$ emission core.}
\begin{tabular}{cccccc} \hline
Emission  &    $A\equiv V \sin(i)$       & $V$ &  I/I$_{\tt c}$      &    FWHM  & $\varphi_0$  \\
component & (\kms)     &    (\kms)         &   &            (phase)                      \\
\hline
 RW Sex   &           & $i = 34\degr$      &       &       &       \\
Narrow (1)&   50.3     & 89.9  &0.19      &  172.1 & 0.00 \\ 
Wide   (2)&  164.6     & 294.4   &0.14      &  484.4 & 0.43\\ 
\hline
\multicolumn{2}{c}{1RXS J064434.5+334451}                        &  $i = 74\degr$    &          \\
Narrow (1)&   68.6     & 71.4  &   0.22      &  216.6  & 0.00 \\ 
Wide   (2)&  297.1     & 309.1  &   0.46      & 1379.9 & 0.42 \\ 
\hline
\end{tabular}
\label{tab:par}
\end{table}

We used Doppler tomography \citep{Marsh:1988aa} to map the H$\alpha$ and H$\beta$
lines using our high-spectral-resolution observations (Fig.~\ref{fig:dopHa}, two upper
lines).
The orbital zero phase
was selected following the  interpretation of \citet{1992A&A...256..433B}
that the narrow  component of H$\alpha$ is a `chromospheric' emission from
the irradiated  hemisphere of the secondary facing the accretion disc.
It corresponds to the strongly concentrated emission inside the Roche lobe
of the secondary star on the Doppler maps.
In addition, the maps show a sign of an extended emission area on the bottom
 of the tomograms. This area corresponds to
the `wide' component of emission lines.

In order to study the structure of the wide emission component,
we fitted the H$\alpha$
profiles with a sum of two Gaussians, denoted here as `narrow' and `wide'.
They are characterized by the peak intensity $I$ and the full width at half-maximum (FWHM),
and their radial velocities depend on the orbital phase
\begin{eqnarray}
v = \gamma - A \times \sin \left[2 \pi (\varphi-\varphi_0) \right],
\end{eqnarray}
where $\gamma$ is the systemic velocity, $A$ is the semi-amplitude of the radial-velocity variation and
$\varphi$ is the orbital phase calculated relative to epoch $T_{0}$ = HJD 245\,7036.83486.
Phase zero $\varphi_0$ is defined by the blue-to-red crossing of the corresponding emission component.
The narrow Gaussian was fitted to the H$\alpha$ profile in individual spectra wherever it was
feasible because at some phases the `narrow' and `wide' components  are indistinguishable.
The mean parameters of both the Gaussian components were deduced as
the average of individual fits and presented in Table~\ref{tab:par}.
We then subtracted the average `narrow' component from the observed profiles and calculated
the Doppler map of the `wide' component (Fig.~\ref{fig:dopHa}, bottom panels).
The tomogram does not show the usual `doughnut' structure commonly observed from
accretion discs. There is also no evidence for the gas stream/disc impact region
emission. Instead, it displays an extended, asymmetric region of emission,
mostly concentrated in the bottom half of the map in its lower-right quadrant.
The brightest area is located far from the position of the WD ($V_x$=0,
$V_y$$\approx$--100\kms),
but does not extend beyond  $\sim$300 \kms\ on the Doppler map.
This is less than the minimal possible Keplerian velocity of the accretion disc with
the largest, tidally truncated radius,
which can be estimated using equation (2.61) from \citet{Warner:1995aa}
\begin{eqnarray}
\label{Eq:truncradius}
   r_{\rm d}({\tt max})  =a  \frac{0.6}{1+q}\,,
\end{eqnarray}
where $a$ is the binary separation and $q$ is the mass ratio.
For the adopted system parameters of RW~Sex ($M_1$=0.84 $M_{\sun}$, $i=34\degr$, $q=0.74$),
the minimal Keplerian velocity of the largest disc
is about 250 \kms, denoted in Doppler maps by the circle.
Because the Keplerian velocity inside of the disc is higher than at the
tidal limitation radius, the accretion disc in Doppler maps must always
be located outside of the circle. However, one can clearly see that the
emission from the `wide' component is concentrated outside the accretion disc.

The tomograms also show another emission region centred at ($V_x$=100 \kms,
$V_y$$\approx$350 \kms). After the removal of the `narrow' component, this
additional, `third' component is visible more clearly in the trailed spectra
and on the Doppler map. It might  be associated with  a low contrast spiral
tidal shock in the outer parts of the accretion disc \citep{Matsuda:1990aa, Steeghs:2001aa,
Kononov:2012aa}.

 \begin{table}
 \centering
    \caption{Parameters of 1RXS J064434.5+334451 from the eclipse modelling. Numbers in parentheses are adopted uncertainties of the fit (see the text). Width$_{spot}$(r$_{out}$) is fixed.}

\begin{tabular}{lclc} \hline
Parameter               &   Value              &   Parameter &   Value     \\   \hline
\multicolumn{2}{c}{System} & \multicolumn{2}{c}{Disc} \\ \cline{1-2}\cline{3-4}
$M_1 (M_{\sun}$)          & 0.73(7)      &             $r_{\tt out}/a$             &         0.33      \\
$M_2 (M_{\sun}$)          & 0.58         &             $h_{\tt out}/r_{\tt out}$       &   0.15(5) \\
$q$                             & 0.80(2)          &           $\gamma_{\tt disc}$       &   1.0(5)      \\
$i$ (\degr)               & 74.0(3)          &             $EXP$      &   0.22(2)  \\ \cline{3-4}
$T_2$ (K)                        & 4250(200)        &                                       & \\
$R_{1}$ (R$_{\sun}$)   & 0.0105       &\multicolumn{2}{c}{Hot spot}              \\
$R_2$ (R$_{\sun}$)    & 0.691       &                                  &  \\ \cline{3-4}
$a$ (R$_{\sun}$)           & 1.92                       & $\phi_{\tt min}$ (\degr)        &   12.4    \\
$\dot{M}$ ($\times10^{-8}M_{\sun}$/yr)  &  0.15--1.5    & $\phi_{\tt max}$ (\degr)        &   42.4(15)\\
distance  (pc)                    & 390--800            & width$_{\tt spot}$ ($r_{\tt out}$)  &    0.05   \\
\hline
\end{tabular}

\label{tab:par2}
\end{table}

We also explored the \ion{Fe}{II} 5316 line (Fig.~\ref{fig:FeII}).  It is centrally placed in
an echelle order and hence secures a better signal-to-noise
ratio ($S/N\approx60$) at the  continuum level than other \ion{Fe}{II}
lines detected in the spectra. In the summed spectrum, the line shows a broad absorption
profile with a relatively strong emission core (Fig.~\ref{fig:FeII}, bottom panel).
Although the line is noisy in individual spectra, it is easily traced in the trailed
spectrum (Fig.~\ref{fig:FeII}, top-left panel), allowing us to construct a decent Doppler
map (Fig.~\ref{fig:FeII}, top-right panel).
The tomogram shows that the emission component emanates from the same source as the `narrow'
emission component of H$\alpha$, which we identify as the heated face of the secondary star.
The absorption component seems to originate in the accretion disc, close to the WD.

\begin{figure}
\setlength{\unitlength}{1mm}
\resizebox{11cm}{!}{
\begin{picture}(100,70)(0,0)
\put (-5,0)    { \includegraphics[width=8.2cm, bb = 100 350 545 745, clip=]{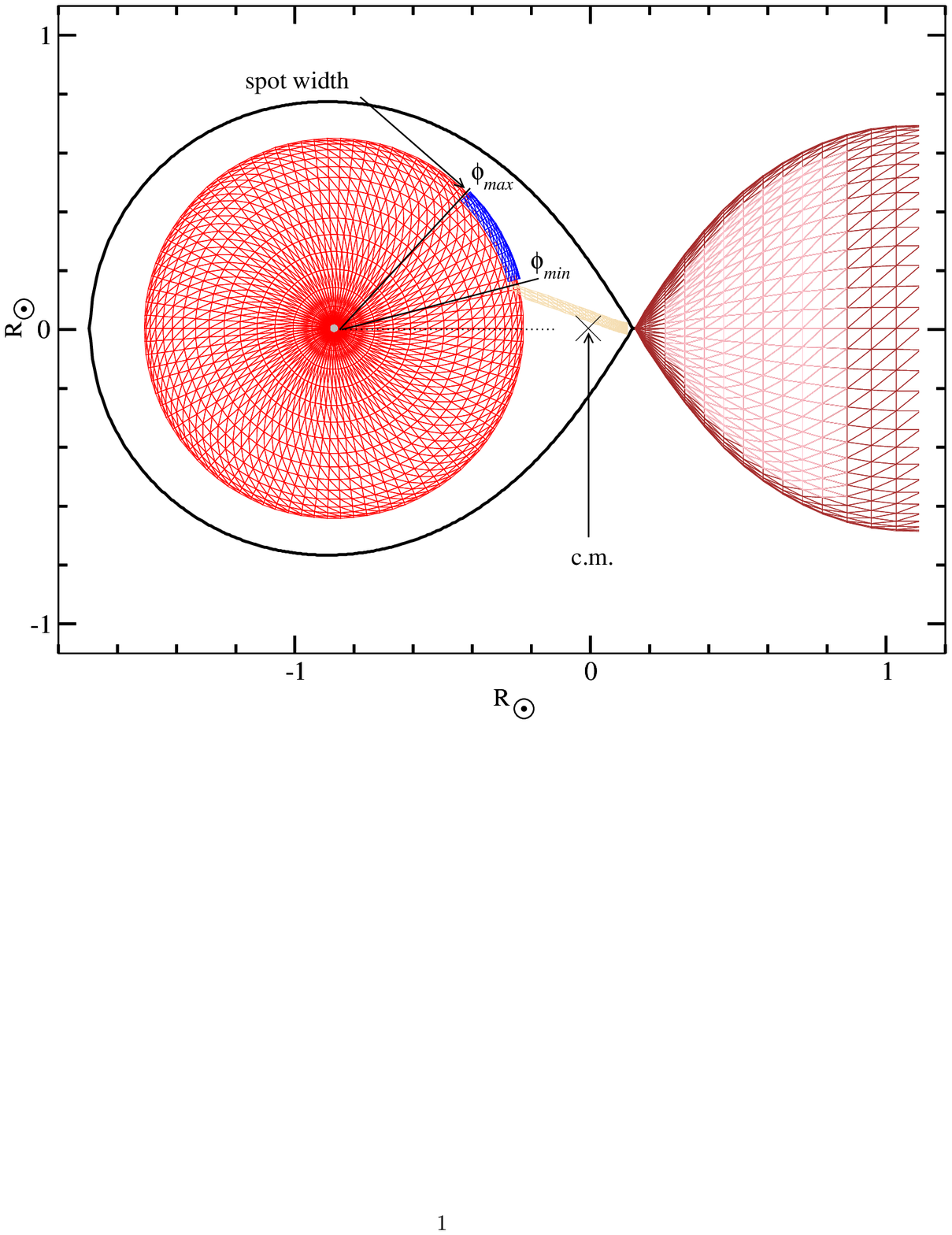}}
\end{picture}}
  \caption{A schematic model of 1RXS~J064434.5+334451.
See description in the text.}
  \label{fig:system}
\end{figure}

\section{Revisiting  1RXS~J064434.5+334451}
 \label{RXSDopmap}

The orbital periods of  1RXS~J064434.5+334451 (0.2694d) and RW~Sex  (0.2451d) are very close and 
the Doppler mapping  analysis of the Balmer lines  
showed remarkably similar structures of  accretion flows in both systems.
An additional advantage of 1RXS~J064434.5+334451, compared with RW~Sex,  is the high inclination of the system with prominent eclipse.
This allows us to   determine exactly the phase 0.0 of the system and make an effort to identify the definite locations of emission-line sources.  
 However,  as we already noted,  the system parameters determined  by \citet{Sing:2007aa}, \citet{Hernandez-Santisteban:2012aa}, \citet{Echevarria:2015aa}, and \citet{Hernandez-Santisteban:2017aa}  for 1RXS~J064434.5+334451 do not agree..  Moreover,  we  note that the system
 parameters of \citet{Hernandez-Santisteban:2012aa} and \citet{Hernandez-Santisteban:2017aa} do not seem realistic because a secondary
 star with a mass of $M_2$$\geq$$0.79M_{\sun}$, as it follows from their measurements, noticeably
 exceeds its Roche lobe for the given orbital period, regardless of whether  it is on the main sequence
 or evolved\footnote{ See the empirical radius-period relationship (2.101) of  \citet{Warner:1995aa}.}.
  This makes it difficult to determine the whereabouts of the sources of emission components.   Therefore, in   subsection~\ref{modelling},  we used another approach  to improve  the system parameters. Unlike the previous attempts, where only the dynamical constrains were  considered,  we rely on the eclipse profile at the same time as radial velocities. We fitted the combined V$-$band light curve of 1RXS~J064434.5+334451 using the same eclipse modelling technique  as in  \citet{Zharikov:2013aa} and \citet{Tovmassian:2014aa}.
 Additionally, in section~\ref{dopmapRSX} we also  reanalyzed  spectral  data from \citet{Hernandez-Santisteban:2012aa,Hernandez-Santisteban:2017aa}  
 with the using newly defined system parameters to determine the location  of the  `wide' emission component source in 1RXS~J064434.5+334451.

   \begin{figure*}
\setlength{\unitlength}{1mm}
\resizebox{11cm}{!}{
\begin{picture}(100,162)(0,0)
\put (-15,0)    { \includegraphics[width=12.5cm, bb = 90 180 480 680, clip=]{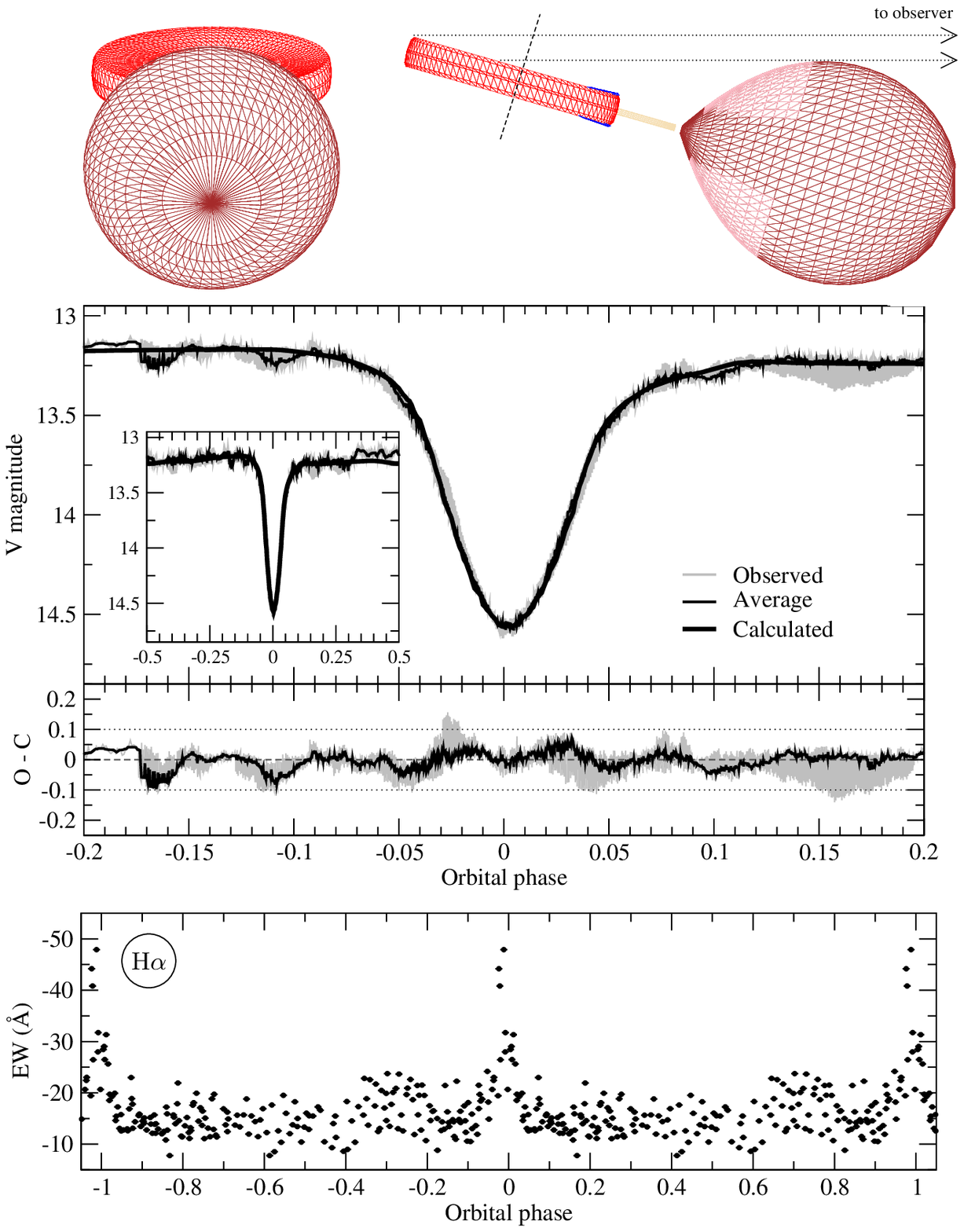}}
\end{picture}}
  \caption{
           Top panel: a projection view of the 1RXS J064434.5+334451 model at orbital phases 0.0 (left) and 0.25 (right).
           Middle panel: the observed (averaged and individual) and best-fitted $V$-band eclipse profiles (top)
           and the corresponding residuals (bottom).
           Bottom panel: EW variation of the H$\alpha$ emission line with orbital phase.}
  \label{fig:phot}
\end{figure*}

   \begin{figure*}
\setlength{\unitlength}{1mm}
\resizebox{11cm}{!}{
\begin{picture}(100,120)(0,0)
\put (-25,0)    { \includegraphics[width=14.5cm, bb = 33 250 560 670, clip=]{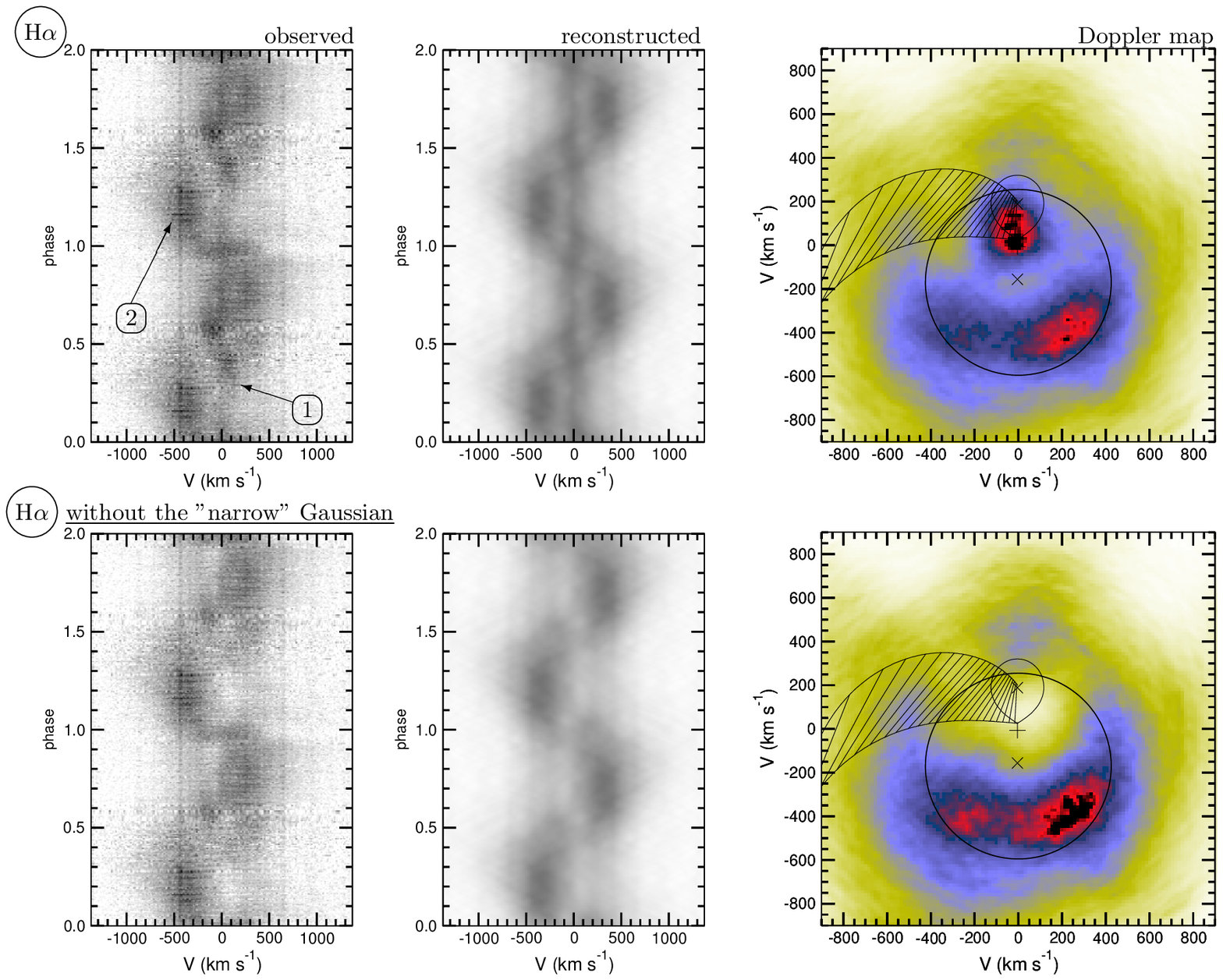}}
\end{picture}}
  \caption{Observed  and  reconstructed  trailed spectra and the corresponding Doppler maps
           (columns from the left to right) of H$\alpha$ of 1RXS~J064434.5+334451.
           The top panels show the observed, unaltered line, while the bottom row shows H$\alpha$
           after the subtraction of the `narrow' component.
           The `narrow' and `wide' components of H$\alpha$ are marked on the observed trailed spectrum
           by '1' and '2', respectively.
            The symbols on the Doppler maps are
            plotted for $M_1=0.73 M_{\sun}$, $i=74\degr$, $q=0.80$.}
  \label{fig:dopHaCV0644}
\end{figure*}

  \subsection{Eclipse modelling and system parameters of 1RXS~J064434.5+334451 }
 \label{modelling}
Light curves of 1RXS J064434.5+334451 in the $V$ band and eclipse profiles in the
$B$, $V$ and $R$ bands were presented by \citet[see their fig.~2]{Sing:2007aa} and
\citet[see their figs~5 and~6]{Hernandez-Santisteban:2017aa}. During the
eclipse, the flux of the object decreases by about $\sim$1.25 mag and
the spectrum slope changes from `blue' to `flat'. However, the Balmer
emission lines do not diminish and they preserve their
one-peaked shape \citep[see their fig.~3]{Sing:2007aa}.

 The modelling technique developed by \citet{Zharikov:2013aa} and \citet{Tovmassian:2014aa} permits us to define parameters of a binary system  from an analysis  of the eclipse  light curve.

Fig.~\ref{fig:system}  shows a geometrical model used for the fitting. It is composed of a concave accretion disc,
a secondary star filling its Roche lobe, a stream from the inner Lagrangian point and a hotspot
area of interaction between the gas stream and the disc, which is located at the outer rim of the
accretion disc.
The concave accretion disc is characterized by the outer radius $r_{\tt out}$, the opening angle $h_{\tt out}$,
as seen from the WD, and the slope $\gamma_{\tt disc}$, controlling the vertical extension of the disc:
  $h = h_{\tt out} (r/r_{\tt out})^{\gamma_{\tt disc}}$.
The outer radius of the accretion disc $r_{\tt out}$
was fixed at the tidal truncation radius $r_d( \tt max)$ (equation \ref{Eq:truncradius}).
We assume that the disc radiates as a blackbody at the local temperature. The radial
distribution across the disc is given by  the equation:
\begin{eqnarray}
T(r) & = &  \left\{\frac{3GM_1\dot{M}}{8\pi\sigma r^3} \times\left(1-
\left[\frac{R_1}{r}\right]^{1/2}\right)\right\}^{EXP},
\label{Tempeq}
\end{eqnarray}
where $M_1$ and $R_1$ are the mass and the radius of the WD, respectively, $\dot{M}$
is the mass-transfer rate, $G$ is the gravitation constant and $\sigma$ is the Stefan--Boltzmann
constant. The radial temperature gradient $EXP$ in the standard accretion disc model is equal to
0.25 \citep[equation 2.35]{Warner:1995aa}, but we allow it to slightly deviate from this value,
following to \citet{2010ApJ...719..271L}.  The temperature of the WD
corresponds to the temperature of the inner edge of the accretion disc $T_1  \equiv
T_{\tt WD} = T_{\tt disc,in}$. We used the mass--radius relationship for WDs from \citet[equation 2.82]{Warner:1995aa}.
The vertical extension and the temperature of the hotspot area linearly decline from maximum values
$h_{\tt s,max}$ and $T_{\tt s,max}$ at the point $\phi_{\tt min}$ to minimum values $h_{\tt s,min} \equiv h_{\tt out}$ and $T_{\tt s,min} \equiv T_{\tt out}$
at $\phi_{\tt min}$, respectively (see Fig.~\ref{fig:system}).

The surface of each system component is divided into a series of triangles, and each
triangle emits as a blackbody with the corresponding temperature. The total flux
from the system is obtained by integrating the emission from all non-occulted
elements lying in the view, and then folded with the response of a photometric
filter. Emission from the accretion stream is not taken into account.
As a result, the code allows us to calculate the light
curve of the binary in a selected photometric band and radial-velocity curves for all
components (e.g., the primary or secondary star, the hot spot, etc).

To fit the eclipse profiles of 1RXS J064434.5+334451, we varied $M_1$, $q$, $\dot{M}$, $EXP$,
$h_{\tt out}$, $\gamma$, the temperature of the secondary $T_2$ and the parameters of the hotspot
with the aim to find the best coincidence
\begin{eqnarray}
\sigma_{\min} =  \sqrt{ \frac{1}{N}\sum^N_{i=1}(O - C)^2}
\end{eqnarray}
between the observed $O$ and calculated $C$ magnitudes in the $[-0.1; 0.1]$ range of orbital
phases. The range was selected arbitrarily in order  to exclude intrinsic disc
variabilities (flickering),  which is not periodic and is stronger  out
of the eclipse. However, we ensured  that the flux  outside  of the $[-0.1; 0.1]$ interval does  not exceed the observed.

 The average $V$-band eclipse profile,  the best  fit and the residuals
 are plotted in  Fig.~\ref{fig:phot}.
  Individual eclipses, which were used to fetch the average profile and
 individual residuals, are also shown as a grey background.  We allowed
 reasonable 10 per cent uncertainties to the fit because the  scatter of
 magnitudes  in the data is about 0.1 mag. The obtained parameters
 of the best fit are given in Table~\ref{tab:par2}. The errors
 were obtained by a variation of each free parameter  while fixing the
 others at their best values.
 The derived system parameters are close to those reported by \citet{Sing:2007aa} and \cite{Echevarria:2015aa}, and they are not
 consistent with those of \citet{Hernandez-Santisteban:2017aa}. 
 
  In particular, the radial-velocity semi-amplitude of the secondary star $K_2 = 191.6$ \kms,
 derived from our model, is in agreement with the value 192.8$\pm$5.6 \kms\ obtained from
 observations \citep{Sing:2007aa}.

  By adopting the best-fitting
 value of $\dot{M}= 0.8\times10^{-8} M_{\sun}$ yr$^{-1}$, we determine the temperature of the outer
 edge of the accretion disc to be about 8500\,K and $\sim$55\,000~K for the inner parts of the disc.
 The hot spot is relatively compact, and its azimuthal extension is about 30$\degr$.
 The highest temperature in the spot is $\sim$10\,500~K, only $\sim$2000\,K hotter
 than the rest of the  disc's edge.  However, the maximal contribution
 of the hotspot to the total flux of the system in the $V$ band is only about
 10 per cent immediately before the eclipse.
 The contribution of the secondary  star is insignificant during all orbital phases.
 Even at the minimum, its flux
 is at least five times lower than of  the
 visible part of the disc.

The radial temperature gradient  $EXP=0.22$ is slightly smaller than the
 standard value 0.25. Models with $EXP=0.25$ always  give a narrower eclipse profile, producing
 deviations not only in the ingress and egress but also in the depth of the eclipse. We
 note that a similar deviation from the standard model was recently found for RW~Sex by
 \citet{2010ApJ...719..271L}.

According to our model, the accretion disc is not totally eclipsed. The
top panel of Fig.~\ref{fig:phot} shows the system from the  observer's
point of view   at the  zero orbital phase. The  stream, the L$_1$ region, the WD and the central part
of the disc around the primary are eclipsed completely. Only external parts of
the accretion  disc and  space above $>0.2r_{\tt out}$ of the WD are
unobscured. The temperature of the visible, inner part of the disc
during the eclipse is below $\sim$20~000K.
While the brightness of the system falls  over $\sim$ three times at the bottom of the eclipse,
 the emission in Balmer lines  remains strong, resulting in an increase
 of equivalent widths  (EWs),   as can be seen from the lower panel of
 Fig.~\ref{fig:phot}.
From the system
geometry and the appearance of the H$\alpha$ line during the eclipse, we
conclude that the H$\alpha$ line forming region is located above or in the outermost
parts, or beyond  the accretion disc.

  \subsection{ Emission-line components and Doppler tomography of 1RXS~J064434.5+334451 }
\label{dopmapRSX}

We used the same method, described above, for the case of RW~Sex, to separate the `narrow' and the `wide' component  of H$\alpha$
emission in 1RXS~J064434.5+334451. The resulting parameters of the fit are presented in Table~\ref{tab:par}.
In Fig.~\ref{fig:dopHaCV0644}, we have reproduced the H$\alpha$ Doppler map
for the  entire line (upper panels) as well as after subtraction of the `narrow' component (bottom panels).
 The `wide' component is concentrated in the lower-right quadrant of the
 tomogram at ($V_x$$\sim$200 \kms, $V_y$$\sim$--350 \kms),  but is extended
 from $V_x$$\sim$--400 to +400 \kms\ and from $V_y$$\sim$--200 to --500\kms.
 These velocities are significantly lower than expected at the tidal truncation radius
 of the disc\footnote{For the adopted system parameters of 1RXS J064434.5+334451, the projected Keplerian velocity at the tidal truncation limit is $\sim450$ km s$^{-1}$.}, and thus the
  location of any emission structures here is unexpected. This could suggest that this
  region is also situated beyond the outer edge of the accretion disc.
  
 Moreover,  the values of radial velocities  in Table~\ref{tab:par} lead to a very important conclusion: the respective radial velocities ($RV_{\tt obs}$) of each component
are a function of the inclination angle   $A \equiv RV_{\tt obs} = V \sin(i)$. Applying the best-estimated inclination angles for each system, we obtain analogous velocities for both objects V$_{\tt narrow} \approx 80$\,km/sec, and V$_{\tt wide} \approx 300$\,km/sec, respectively. The slightly spread of those values can be owned by the differences in component masses in the systems and ambiguity of  inclination angle for RW~Sex. 
This strongly supports the idea that the `wide' component belongs to the orbital plane and does not originate in a wind perpendicular to the accretion disc.

  \section{Accretion flow structure in long-period NL systems}

It is widely accepted that NLs in a high state have a high mass-transfer rate.
As a result, they have hot ($\ga 10\,000$~K), opticallythick accretion discs in an almost steady
state, which are expected to produce
broad absorption lines instead of emission lines, which is common among other CVs in quiescence.
Nevertheless, we routinely observe emission features in many NLs. Usually, these emissions lines
appear simply single-peaked in low-resolution spectra.
However, the Balmer line profiles in the high-resolution spectra of two NL systems, RW~Sex and
1RXS~J064434.5+334451, are complex and comprised of an absorption component from the optically
thick accretion disc, and of at least two
 emission components, labelled here as `narrow' and `wide'.
The  `narrow'  component with a low radial-velocity amplitude
originates at L$_1$ and/or the irradiated surface of the secondary facing the
disc. Meanwhile, the origin of the `wide' component is not clear.
We summarize its main features as follows.
\begin{enumerate}
\item The radial velocity  varies with the orbital period.
\item The orbital phase is shifted relative to the `narrow' component by about
0.43 (rotated clockwise on the Doppler maps).
\item Unlike the \ion{He}{II} line, the `wide' component  
is not related to the WD
      \citep[see the \ion{He}{II} tomogram of 1RXS~J064434.5+334451 in][]{Hernandez-Santisteban:2017aa}.
\item The wide component does not disappear during  eclipses.
\item The spot created by the wide component  on the Doppler maps appears to be identical  for both the low- and the high-inclination systems.
\item The shape and  large size of that spot on the Doppler maps indicates strong
velocity dispersion of the forming region.
\end{enumerate}

In order to examine how unique such a structure  is, we visually inspected the available Doppler
maps and/or trailed spectra of non-magnetic NL systems with orbital periods longer than 4~h.
Unfortunately, although the current edition of the \citet{RitterKolb} catalogue (update 7.24, 2016)
lists more than 35 such NLs, only about a quarter of them were studied spectroscopically. However,
we found that most of them are high-inclination eclipsing systems exhibiting, similarly to
1RXS~J064434.5+334451, single-peaked Balmer, \ion{He}{I} and occasionally \ion{He}{II} emission
lines in and out of eclipses (see, e.g., RW~Tri, IX~Vel,  V347~Pup, V3885~Sgr, AC~Cnc, V363~Aur, BF~Eri --
\citealt{Kaitchuck:1983aa,Kubiak:1999aa,Thoroughgood:2004aa,Thoroughgood:2005aa,Hartley:2005aa,Neustroev:2008aa}).
Moreover, their Doppler maps and/or trailed spectra also show similarities
to those presented in this paper, which allows us to propose a common behaviour of an accretion flow in long-period NLs.

In order to explain the presence of single-peaked emission line profiles in high-inclination CVs,
several possible models have been suggested,
such as Stark broadening \citep{Lin:1988aa}, magnetic
accretion \citep{Williams:1989aa}, wind emission \citep{Honeycutt:1986aa, Murray:1996aa, Matthews:2015aa}, disc-overflow
accretion \citep{Hellier:1994aa}, and  an extended bright spot as the dominant source of emission lines
\citep{Dhillon:1997aa, Tovmassian:2014aa}.

The first three are related to the WD and/or the
innermost part of the accretion disc. It is naturally expected that a source  of emission lines in those
models will demonstrate an orbital motion, which is related to the position of the WD and which
we do not observe in Balmer lines of RW~Sex and 1RXS J064434.5+334451.
In fact, a perpendicular wind in a high-inclination system such as the latter will produce an 
emission line with practically zero velocity.
This phenomenon was observed during the super-outburst of V455\,And \citep{2011IAUS..275..311T}. 
The disc during the outburst becomes optically thick,  similar to NLs. Meanwhile,  emission lines 
are produced in the wind, converting a perfect ring  depicting quiescent accretion disc 
into a concentrated spot at the centre of Doppler maps throughout the super-outburst. 
In the case of  RW~Sex and 1RXS~J064434.5+334451, we observe quite a different effect -the higher the inclination angle 
of the binary plane, the higher the velocity of the `wide' component. Therefore, the matter, emitting the `wide' component,  
 is confined to the orbital plane, but not within the tidal truncation radius of the accretion disc.

The disc-overflow model predicts
that a significant fraction of the stream can overflow the outer disc edge and hit the far, following
side of the disc  close to its circularization radius \citep{DiscOverflowModel}.
Neither the velocity coordinates nor the azimuthal coordinates of the `wide' emission components are
consistent with this prediction.  The model of an extended bright spot was developed to explain
unusual properties of the SW~Sex stars. In this respect, we note that \citet{Hernandez-Santisteban:2017aa}
suggested that the characteristics of 1RXS~J064434.5+334451 are consistent with an SW~Sex star. However,
although some similarities can indeed be found, we doubt this classification.

One of the defining properties of SW~Sex objects is that in the eclipsing stars the emission-line radial-velocity curves show a substantial delay ($\sim$0.2 orbital cycle) with respect to the motion of the WD.
Such a behaviour can be understood by assuming that the primary source of emission lines in the
SW~Sex objects is the hotspot \citep{Tovmassian:2014aa}. However, the contribution of the hotspot is
negligible in both the objects of this study and the phasing of the radial-velocity curve of
1RXS~J064434.5+334451 agrees well with that expected from the eclipses (see fig.~9 in
\citealt{Hernandez-Santisteban:2017aa}). 
Also, the high-velocity (broad) emission components in both objects show maximum blueshift near phase
0.25 (Figs.~\ref{fig:dopHa} and \ref{fig:dopHaCV0644}), whereas in the confirmed SW~Sex stars maximum
blueshift of high-velocity emission S-waves is observed near phase $\sim$0.5 (see e.g. fig.~3 in \citealt{Dhillon:2013aa}
and fig.~6 in \citealt{Tovmassian:2014aa}).
We believe that these differences are essential, and thus the
suggestion of \citet{Hernandez-Santisteban:2017aa} that 1RXS~J064434.5+334451 is an SW~Sex star is not valid.

Thus, neither of the models described above can explain the emission structure of RW~Sex and
1RXS~J064434.5+334451. However,  \citet{Bisikalo98} pointed out that in high-mass-transfer rate
binaries, the matter can escape the accretion disc and create a halo around the disc.
According to their three-dimensional gas-dynamical simulations of accretion flows in a 5.4-h system
with typical stellar parameters and a mass-transfer rate of $10^{-8}$ M$_{\sun}$/year, an extended
low-velocity region (hereafter called an `outflow zone') is filled with matter pouring from the
disc \citep{Bisikalo:2008aa, Kononov:2012aa}. This region is located practically on the opposite
side of the disc  with respect to the hotspot (labelled as A and B in fig.~3 in
\citealt{Bisikalo:2008aa}).  In the synthetic Doppler map, the  position of the region B corresponds
exactly to the large spot  created by the `wide' component of emission lines (Figs~\ref{fig:dopHa} and
\ref{fig:dopHaCV0644}).  The material in the outflow zone has a substantial velocity dispersion and hence
does not form a concentrated spot.  The outflow zone is not totally eclipsed even in a high-inclination system.
Also,  the material from this zone comes to  form a circumbinary ring  \citep{Bisikalo:2009aa, Bisikalo:2010aa},
evidence of which was found in some NLs (including RW~Sex) in far-infrared observations \citep{Hoard:2014aa}.
Therefore,  we suppose this region (the zone of outflow) as a possible source of
the `wide' component of the Balmer line profiles in RW~Sex and 1RXS~J064434.5+334451, and possibly other
long-period NLs. 
The same region was also recently proposed by  \citet{Tovmassian:2014aa} to be responsible for the appearance
of absorption dips in the emission lines of SW~Sex stars around phase 0.4--0.7.

The difference in the appearance of this region in SW~Sex stars, the majority of which are found 
with orbital periods shorter than 4~h\footnote{See D. W. Hoard's Big List of SW Sextantis Stars \citep{Hoard2003}: available at http://www.dwhoard.com/biglist.}, and long-period NLs is intriguing.
We believe that the key is the mass-transfer rate. Indeed, according to standard evolutionary theory,
the mass-transfer rate decreases during the CV evolution. \citet{Howell:2001} showed, through population 
synthesis, that the mass-transfer rate in CVs with orbital periods of 3--4~h is about 5--10 times lower 
than in CVs with periods longer than $\sim$6~h. 
The immediate consequence  is a substantially different temperature of the accretion disc
in the longer and  shorter orbital period systems (see e.g. equation~\ref{Tempeq}). Below, we give our
qualitative view on the observational transformation of NLs during their evolution from longer
to shorter orbital periods.

\begin{description}
\item In long-period systems with a high mass-transfer rate, the disc is hot, dense
and optically thick. The outer edge of the disc reaches $\sog$10\,000~K and emits mainly in the continuum,
producing broad absorption lines, like the atmosphere of a B- or A-type star. The
hotspot contributes only modestly to the continuum emission because its material is dense, optically
thick and the difference in temperatures between the disc edge and the hotspot is relatively small. The emission
lines are primarily formed outside of the accretion disc, in the outflow zone (`wide' component) and on the
surface of the secondary (`narrow' component). The objects observationally resemble
RW Sex. \\

\item At shorter orbital periods, the mass-transfer rate diminishes and the temperature
at the disc edge also drops below 10\,000~K, while the disc generally remains hot and dense (steady-state regime). Less matter leaves the disc and the temperature at the outflow zone is much lower.
Hence, the outflow zone ceases to produce emission lines. Instead, it manifests itself as
transient absorption features at orbital phases around $\sim$0.5. 
The irradiation of the secondary is probably not sufficient enough to contribute
to emission lines. Meanwhile, the difference in temperatures between the disc edge and the hotspot increases
and the latter becomes the dominant source of emission. Such an object is observed as an SW~Sex star.
\end{description}

\section{Conclusions}
\label{conclud}
With the help of  high-resolution spectroscopy, we clearly showed that the H$\alpha$ emission-line
profiles in the spectra of two long-period ($\sim$6~h) NL systems, RW~Sex and 1RXS~J064434.5+334451,
have a complex structure consisting of at least two distinct, variable components.
These emission components (`narrow' and `wide') create identical structures in the Doppler maps
of both objects. We also found hints of similar emission structures in other long-period NLs.
The source of the narrow, low-velocity component is located  on the surface of the secondary
star facing the accretion disc, or near the $L_1$ point. The other, wide component is probably
not related to the WD or the central parts of the accretion disc but  emanates from the outer
side of the disc. We  propose that its source is an extended, low-velocity region in the outskirts
of the accretion disc, on the opposite side with respect to the hotspot.
This naturally explains the profiles of Balmer emission lines observed in
long orbital period NLs in which both components can hardly be separated in low-resolution spectra.

\section*{Acknowledgments}
 This work is based upon observations carried out at the OAN SPM, Baja California, M\'{e}xico. 
GT and SZ acknowledge PAPIIT grants IN108316, IN100617, and CONACyT grant 166376.
 We are grateful to Dr.  J.~V. Hern\'{a}ndez Santisteban for the opportunity to work
with the original data of 1RXS~J064434.5+334451. We thank the daytime and night support staff at the OAN-SPM for facilitating and helping obtain our observations.

\bibliographystyle{mnras}
\bibliography{rwsex}

\begin{thebibliography}{}
\makeatletter
\relax
\def\mn@urlcharsother{\let\do\@makeother \do\$\do\&\do\#\do\^\do\_\do\%\do\~}
\def\mn@doi{\begingroup\mn@urlcharsother \@ifnextchar [ {\mn@doi@}
  {\mn@doi@[]}}
\def\mn@doi@[#1]#2{\def\@tempa{#1}\ifx\@tempa\@empty \href
  {http://dx.doi.org/#2} {doi:#2}\else \href {http://dx.doi.org/#2} {#1}\fi
  \endgroup}
\def\mn@eprint#1#2{\mn@eprint@#1:#2::\@nil}
\def\mn@eprint@arXiv#1{\href {http://arxiv.org/abs/#1} {{\tt arXiv:#1}}}
\def\mn@eprint@dblp#1{\href {http://dblp.uni-trier.de/rec/bibtex/#1.xml}
  {dblp:#1}}
\def\mn@eprint@#1:#2:#3:#4\@nil{\def\@tempa {#1}\def\@tempb {#2}\def\@tempc
  {#3}\ifx \@tempc \@empty \let \@tempc \@tempb \let \@tempb \@tempa \fi \ifx
  \@tempb \@empty \def\@tempb {arXiv}\fi \@ifundefined
  {mn@eprint@\@tempb}{\@tempb:\@tempc}{\expandafter \expandafter \csname
  mn@eprint@\@tempb\endcsname \expandafter{\@tempc}}}

\bibitem[\protect\citeauthoryear{{Beuermann}, {Stasiewski}  \&
  {Schwope}}{{Beuermann} et~al.}{1992}]{1992A&A...256..433B}
{Beuermann} K.,  {Stasiewski} U.,   {Schwope} A.~D.,  1992, \aap, \href
  {http://adsabs.harvard.edu/abs/1992A%26A...256..433B} {256, 433}

\bibitem[\protect\citeauthoryear{{Bisikalo}}{{Bisikalo}}{2009}]{Bisikalo:2009aa}
{Bisikalo} D.~V.,  2009, in {Zhelyazkov} I.,  ed.,  American Institute of
  Physics Conference Series Vol. 1121, American Institute of Physics Conference
  Series. pp 3--10, \mn@doi{10.1063/1.3137942}

\bibitem[\protect\citeauthoryear{{Bisikalo}}{{Bisikalo}}{2010}]{Bisikalo:2010aa}
{Bisikalo} D.,  2010, in {Pr{\v s}a} A.,  {Zejda} M.,  eds,  Astronomical
  Society of the Pacific Conference Series Vol. 435, Binaries - Key to
  Comprehension of the Universe. p.~287

\bibitem[\protect\citeauthoryear{{Bisikalo}, {Boyarchuk}, {Chechetkin},
  {Kuznetsov}  \& {Molteni}}{{Bisikalo} et~al.}{1998}]{Bisikalo98}
{Bisikalo} D.~V.,  {Boyarchuk} A.~A.,  {Chechetkin} V.~M.,  {Kuznetsov} O.~A.,
   {Molteni} D.,  1998, \mn@doi [\mnras] {10.1046/j.1365-8711.1998.01815.x},
  \href {http://adsabs.harvard.edu/abs/1998MNRAS.300...39B} {300, 39}

\bibitem[\protect\citeauthoryear{{Bisikalo}, {Kononov}, {Kaigorodov}, {Zhilkin}
   \& {Boyarchuk}}{{Bisikalo} et~al.}{2008}]{Bisikalo:2008aa}
{Bisikalo} D.~V.,  {Kononov} D.~A.,  {Kaigorodov} P.~V.,  {Zhilkin} A.~G.,
  {Boyarchuk} A.~A.,  2008, \mn@doi [Astronomy Reports]
  {10.1134/S1063772908040069}, \href
  {http://adsabs.harvard.edu/abs/2008ARep...52..318B} {52, 318}

\bibitem[\protect\citeauthoryear{{Coppejans}, {K{\"o}rding}, {Miller-Jones},
  {Rupen}, {Knigge}, {Sivakoff}  \& {Groot}}{{Coppejans}
  et~al.}{2015}]{Coppejans:2015aa}
{Coppejans} D.~L.,  {K{\"o}rding} E.~G.,  {Miller-Jones} J.~C.~A.,  {Rupen}
  M.~P.,  {Knigge} C.,  {Sivakoff} G.~R.,   {Groot} P.~J.,  2015, \mn@doi
  [\mnras] {10.1093/mnras/stv1225}, \href
  {http://adsabs.harvard.edu/abs/2015MNRAS.451.3801C} {451, 3801}

\bibitem[\protect\citeauthoryear{{Dhillon}, {Marsh}  \& {Jones}}{{Dhillon}
  et~al.}{1997}]{Dhillon:1997aa}
{Dhillon} V.~S.,  {Marsh} T.~R.,   {Jones} D.~H.~P.,  1997, \mn@doi [\mnras]
  {10.1093/mnras/291.4.694}, \href
  {http://adsabs.harvard.edu/abs/1997MNRAS.291..694D} {291, 694}

\bibitem[\protect\citeauthoryear{{Dhillon}, {Smith}  \& {Marsh}}{{Dhillon}
  et~al.}{2013}]{Dhillon:2013aa}
{Dhillon} V.~S.,  {Smith} D.~A.,   {Marsh} T.~R.,  2013, \mn@doi [\mnras]
  {10.1093/mnras/sts294}, \href
  {http://adsabs.harvard.edu/abs/2013MNRAS.428.3559D} {428, 3559}

\bibitem[\protect\citeauthoryear{{Echevarria}}{{Echevarria}}{2015}]{Echevarria:2015aa}
{Echevarria} J.,  2015, IAU General Assembly, \href
  {http://adsabs.harvard.edu/abs/2015IAUGA..2258553E} {22, 2258553}

\bibitem[\protect\citeauthoryear{{Greenstein} \& {Oke}}{{Greenstein} \&
  {Oke}}{1982}]{1982ApJ...258..209G}
{Greenstein} J.~L.,  {Oke} J.~B.,  1982, \mn@doi [\apj] {10.1086/160069}, \href
  {http://adsabs.harvard.edu/abs/1982ApJ...258..209G} {258, 209}

\bibitem[\protect\citeauthoryear{{Hartley}, {Murray}, {Drew}  \&
  {Long}}{{Hartley} et~al.}{2005}]{Hartley:2005aa}
{Hartley} L.~E.,  {Murray} J.~R.,  {Drew} J.~E.,   {Long} K.~S.,  2005, \mn@doi
  [\mnras] {10.1111/j.1365-2966.2005.09447.x}, \href
  {http://adsabs.harvard.edu/abs/2005MNRAS.363..285H} {363, 285}

\bibitem[\protect\citeauthoryear{{Hellier} \& {Robinson}}{{Hellier} \&
  {Robinson}}{1994}]{Hellier:1994aa}
{Hellier} C.,  {Robinson} E.~L.,  1994, \mn@doi [\apjl] {10.1086/187484}, \href
  {http://adsabs.harvard.edu/abs/1994ApJ...431L.107H} {431, L107}

\bibitem[\protect\citeauthoryear{{Hern{\'a}ndez Santisteban}}{{Hern{\'a}ndez
  Santisteban}}{2012}]{Hernandez-Santisteban:2012aa}
{Hern{\'a}ndez Santisteban} J.~V.,  2012, \memsai, \href
  {http://adsabs.harvard.edu/abs/2012MmSAI..83..729H} {83, 729}

\bibitem[\protect\citeauthoryear{{Hern{\'a}ndez Santisteban},
  {Echevarr{\'{\i}}a}, {Michel}  \& {Costero}}{{Hern{\'a}ndez Santisteban}
  et~al.}{2017}]{Hernandez-Santisteban:2017aa}
{Hern{\'a}ndez Santisteban} J.~V.,  {Echevarr{\'{\i}}a} J.,  {Michel} R.,
  {Costero} R.,  2017, \mn@doi [\mnras] {10.1093/mnras/stw2282}, \href
  {http://adsabs.harvard.edu/abs/2017MNRAS.464..104H} {464, 104}

\bibitem[\protect\citeauthoryear{{Hoard}, {Szkody}, {Froning}, {Long}  \&
  {Knigge}}{{Hoard} et~al.}{2003}]{Hoard2003}
{Hoard} D.~W.,  {Szkody} P.,  {Froning} C.~S.,  {Long} K.~S.,   {Knigge} C.,
  2003, \mn@doi [\aj] {10.1086/378605}, \href
  {http://adsabs.harvard.edu/abs/2003AJ....126.2473H} {126, 2473}

\bibitem[\protect\citeauthoryear{{Hoard} et~al.,}{{Hoard}
  et~al.}{2014}]{Hoard:2014aa}
{Hoard} D.~W.,  et~al., 2014, \mn@doi [\apj] {10.1088/0004-637X/786/1/68},
  \href {http://adsabs.harvard.edu/abs/2014ApJ...786...68H} {786, 68}

\bibitem[\protect\citeauthoryear{{Honeycutt}, {Schlegel}  \&
  {Kaitchuck}}{{Honeycutt} et~al.}{1986}]{Honeycutt:1986aa}
{Honeycutt} R.~K.,  {Schlegel} E.~M.,   {Kaitchuck} R.~H.,  1986, \mn@doi
  [\apj] {10.1086/163997}, \href
  {http://adsabs.harvard.edu/abs/1986ApJ...302..388H} {302, 388}

\bibitem[\protect\citeauthoryear{{Howell}, {Nelson}  \& {Rappaport}}{{Howell}
  et~al.}{2001}]{Howell:2001}
{Howell} S.~B.,  {Nelson} L.~A.,   {Rappaport} S.,  2001, \mn@doi [\apj]
  {10.1086/319776}, \href {http://adsabs.harvard.edu/abs/2001ApJ...550..897H}
  {550, 897}

\bibitem[\protect\citeauthoryear{{Kaitchuck}, {Schlegel}  \&
  {Honeycutt}}{{Kaitchuck} et~al.}{1983}]{Kaitchuck:1983aa}
{Kaitchuck} R.~H.,  {Schlegel} E.~M.,   {Honeycutt} R.~K.,  1983, \mn@doi
  [\apj] {10.1086/160863}, \href
  {http://adsabs.harvard.edu/abs/1983ApJ...267..239K} {267, 239}

\bibitem[\protect\citeauthoryear{{Kononov}, {Giovannelli}, {Bruni}  \&
  {Bisikalo}}{{Kononov} et~al.}{2012}]{Kononov:2012aa}
{Kononov} D.~A.,  {Giovannelli} F.,  {Bruni} I.,   {Bisikalo} D.~V.,  2012,
  \mn@doi [\aap] {10.1051/0004-6361/201016334}, \href
  {http://adsabs.harvard.edu/abs/2012A%26A...538A..94K} {538, A94}

\bibitem[\protect\citeauthoryear{{Kubiak}, {Pojmanski}  \&
  {Krzeminski}}{{Kubiak} et~al.}{1999}]{Kubiak:1999aa}
{Kubiak} M.,  {Pojmanski} G.,   {Krzeminski} W.,  1999, \actaa, \href
  {http://adsabs.harvard.edu/abs/1999AcA....49...73K} {49, 73}

\bibitem[\protect\citeauthoryear{{Kunze}, {Speith}  \& {Hessman}}{{Kunze}
  et~al.}{2001}]{DiscOverflowModel}
{Kunze} S.,  {Speith} R.,   {Hessman} F.~V.,  2001, \mn@doi [\mnras]
  {10.1046/j.1365-8711.2001.04057.x}, \href
  {http://adsabs.harvard.edu/abs/2001MNRAS.322..499K} {322, 499}

\bibitem[\protect\citeauthoryear{{Levine} \& {Chakarabarty}}{{Levine} \&
  {Chakarabarty}}{1995}]{Levine1995}
{Levine} S.,  {Chakarabarty} D.,  1995, IA-UNAM Technical Report MU-94-04

\bibitem[\protect\citeauthoryear{{Lin}, {Williams}  \& {Stover}}{{Lin}
  et~al.}{1988}]{Lin:1988aa}
{Lin} D.~N.~C.,  {Williams} R.~E.,   {Stover} R.~J.,  1988, \mn@doi [\apj]
  {10.1086/166185}, \href {http://adsabs.harvard.edu/abs/1988ApJ...327..234L}
  {327, 234}

\bibitem[\protect\citeauthoryear{{Linnell}, {Godon}, {Hubeny}, {Sion}  \&
  {Szkody}}{{Linnell} et~al.}{2010}]{2010ApJ...719..271L}
{Linnell} A.~P.,  {Godon} P.,  {Hubeny} I.,  {Sion} E.~M.,   {Szkody} P.,
  2010, \apj, 719, 271

\bibitem[\protect\citeauthoryear{{Lynden-Bell}}{{Lynden-Bell}}{1969}]{Lynden-Bell:1969aa}
{Lynden-Bell} D.,  1969, \mn@doi [\nat] {10.1038/223690a0}, \href
  {http://adsabs.harvard.edu/abs/1969Natur.223..690L} {223, 690}

\bibitem[\protect\citeauthoryear{{Marsh} \& {Horne}}{{Marsh} \&
  {Horne}}{1988}]{Marsh:1988aa}
{Marsh} T.~R.,  {Horne} K.,  1988, \mnras, \href
  {http://adsabs.harvard.edu/abs/1988MNRAS.235..269M} {235, 269}

\bibitem[\protect\citeauthoryear{{Matsuda}, {Sekino}, {Shima}, {Sawada}  \&
  {Spruit}}{{Matsuda} et~al.}{1990}]{Matsuda:1990aa}
{Matsuda} T.,  {Sekino} N.,  {Shima} E.,  {Sawada} K.,   {Spruit} H.,  1990,
  \aap, \href {http://adsabs.harvard.edu/abs/1990A%26A...235..211M} {235, 211}

\bibitem[\protect\citeauthoryear{{Matthews}, {Knigge}, {Long}, {Sim}  \&
  {Higginbottom}}{{Matthews} et~al.}{2015}]{Matthews:2015aa}
{Matthews} J.~H.,  {Knigge} C.,  {Long} K.~S.,  {Sim} S.~A.,   {Higginbottom}
  N.,  2015, \mn@doi [\mnras] {10.1093/mnras/stv867}, \href
  {http://adsabs.harvard.edu/abs/2015MNRAS.450.3331M} {450, 3331}

\bibitem[\protect\citeauthoryear{{Murray} \& {Chiang}}{{Murray} \&
  {Chiang}}{1996}]{Murray:1996aa}
{Murray} N.,  {Chiang} J.,  1996, \mn@doi [\nat] {10.1038/382789a0}, \href
  {http://adsabs.harvard.edu/abs/1996Natur.382..789M} {382, 789}

\bibitem[\protect\citeauthoryear{{Neustroev} \& {Zharikov}}{{Neustroev} \&
  {Zharikov}}{2008}]{Neustroev:2008aa}
{Neustroev} V.~V.,  {Zharikov} S.,  2008, \mn@doi [\mnras]
  {10.1111/j.1365-2966.2008.12930.x}, \href
  {http://adsabs.harvard.edu/abs/2008MNRAS.386.1366N} {386, 1366}

\bibitem[\protect\citeauthoryear{Neustroev et~al.,}{Neustroev
  et~al.}{2017}]{NeustroevSSS}
Neustroev V.~V.,  et~al., 2017, \mn@doi [\mnras] {10.1093/mnras/stx084}, 467,
  597

\bibitem[\protect\citeauthoryear{{Noebauer}, {Long}, {Sim}  \&
  {Knigge}}{{Noebauer} et~al.}{2010}]{Noebauer:2010aa}
{Noebauer} U.~M.,  {Long} K.~S.,  {Sim} S.~A.,   {Knigge} C.,  2010, \mn@doi
  [\apj] {10.1088/0004-637X/719/2/1932}, \href
  {http://adsabs.harvard.edu/abs/2010ApJ...719.1932N} {719, 1932}

\bibitem[\protect\citeauthoryear{{Perryman}, {Lindegren}, {Kovalevsky}, {Hoeg}
  \& {Bastian}}{{Perryman} et~al.}{1997}]{Perryman:1997aa}
{Perryman} M.~A.~C.,  {Lindegren} L.,  {Kovalevsky} J.,  {Hoeg} E.,   {Bastian}
  U.,  1997, \aap, \href {http://adsabs.harvard.edu/abs/1997A%26A...323L..49P}
  {323, L49}

\bibitem[\protect\citeauthoryear{{Prinja}, {Long}, {Froning}, {Knigge},
  {Witherick}, {Clark}  \& {Ringwald}}{{Prinja}
  et~al.}{2003}]{2003MNRAS.340..551P}
{Prinja} R.~K.,  {Long} K.~S.,  {Froning} C.~S.,  {Knigge} C.,  {Witherick}
  D.~K.,  {Clark} J.~S.,   {Ringwald} F.~A.,  2003, \mn@doi [\mnras]
  {10.1046/j.1365-8711.2003.06307.x}, \href
  {http://adsabs.harvard.edu/abs/2003MNRAS.340..551P} {340, 551}

\bibitem[\protect\citeauthoryear{{Ritter} \& {Kolb}}{{Ritter} \&
  {Kolb}}{2003}]{RitterKolb}
{Ritter} H.,  {Kolb} U.,  2003, \mn@doi [\aap] {10.1051/0004-6361:20030330},
  \href {http://adsabs.harvard.edu/abs/2003A%26A...404..301R} {404, 301}

\bibitem[\protect\citeauthoryear{{Sing}, {Green}, {Howell}, {Holberg},
  {Lopez-Morales}, {Shaw}  \& {Schmidt}}{{Sing} et~al.}{2007}]{Sing:2007aa}
{Sing} D.~K.,  {Green} E.~M.,  {Howell} S.~B.,  {Holberg} J.~B.,
  {Lopez-Morales} M.,  {Shaw} J.~S.,   {Schmidt} G.~D.,  2007, \mn@doi [\aap]
  {10.1051/0004-6361:20078026}, \href
  {http://adsabs.harvard.edu/abs/2007A%26A...474..951S} {474, 951}

\bibitem[\protect\citeauthoryear{{Steeghs}}{{Steeghs}}{2001}]{Steeghs:2001aa}
{Steeghs} D.,  2001, in {Boffin} H.~M.~J.,  {Steeghs} D.,   {Cuypers} J.,  eds,
   Lecture Notes in Physics, Berlin Springer Verlag Vol. 573, Astrotomography,
  Indirect Imaging Methods in Observational Astronomy. p.~45 (\mn@eprint {}
  {astro-ph/0012353})

\bibitem[\protect\citeauthoryear{{Thoroughgood}, {Dhillon}, {Watson},
  {Buckley}, {Steeghs}  \& {Stevenson}}{{Thoroughgood}
  et~al.}{2004}]{Thoroughgood:2004aa}
{Thoroughgood} T.~D.,  {Dhillon} V.~S.,  {Watson} C.~A.,  {Buckley} D.~A.~H.,
  {Steeghs} D.,   {Stevenson} M.~J.,  2004, \mn@doi [\mnras]
  {10.1111/j.1365-2966.2004.08135.x}, \href
  {http://adsabs.harvard.edu/abs/2004MNRAS.353.1135T} {353, 1135}

\bibitem[\protect\citeauthoryear{{Thoroughgood} et~al.,}{{Thoroughgood}
  et~al.}{2005}]{Thoroughgood:2005aa}
{Thoroughgood} T.~D.,  et~al., 2005, \mn@doi [\mnras]
  {10.1111/j.1365-2966.2004.08613.x}, \href
  {http://adsabs.harvard.edu/abs/2005MNRAS.357..881T} {357, 881}

\bibitem[\protect\citeauthoryear{{Thorstensen}, {Davis}  \&
  {Ringwald}}{{Thorstensen} et~al.}{1991}]{Thorstensen:1991aa}
{Thorstensen} J.~R.,  {Davis} M.~K.,   {Ringwald} F.~A.,  1991, \mn@doi [\aj]
  {10.1086/115902}, \href {http://adsabs.harvard.edu/abs/1991AJ....102..683T}
  {102, 683}

\bibitem[\protect\citeauthoryear{{Tovmassian}, {G{\"a}nsicke}, {Zharikov},
  {Ramirez}  \& {Diaz}}{{Tovmassian} et~al.}{2011}]{2011IAUS..275..311T}
{Tovmassian} G.,  {G{\"a}nsicke} B.,  {Zharikov} S.,  {Ramirez} A.,   {Diaz}
  M.,  2011, in {Romero} G.~E.,  {Sunyaev} R.~A.,   {Belloni} T.,  eds,  IAU
  Symposium Vol. 275, Jets at All Scales. pp 311--312,
  \mn@doi{10.1017/S174392131001625X}

\bibitem[\protect\citeauthoryear{{Tovmassian}, {Stephania Hernandez},
  {Gonz{\'a}lez-Buitrago}, {Zharikov}  \&
  {Garc{\'{\i}}a-D{\'{\i}}az}}{{Tovmassian} et~al.}{2014}]{Tovmassian:2014aa}
{Tovmassian} G.,  {Stephania Hernandez} M.,  {Gonz{\'a}lez-Buitrago} D.,
  {Zharikov} S.,   {Garc{\'{\i}}a-D{\'{\i}}az} M.~T.,  2014, \mn@doi [\aj]
  {10.1088/0004-6256/147/3/68}, \href
  {http://adsabs.harvard.edu/abs/2014AJ....147...68T} {147, 68}

\bibitem[\protect\citeauthoryear{{Vitello} \& {Shlosman}}{{Vitello} \&
  {Shlosman}}{1993}]{Vitello:1993aa}
{Vitello} P.,  {Shlosman} I.,  1993, \mn@doi [\apj] {10.1086/172799}, \href
  {http://adsabs.harvard.edu/abs/1993ApJ...410..815V} {410, 815}

\bibitem[\protect\citeauthoryear{{Warner}}{{Warner}}{1995}]{Warner:1995aa}
{Warner} B.,  1995, Cambridge Astrophysics Series, \href
  {http://adsabs.harvard.edu/abs/1995CAS....28.....W} {28}

\bibitem[\protect\citeauthoryear{{Williams}}{{Williams}}{1989}]{Williams:1989aa}
{Williams} R.~E.,  1989, \mn@doi [\aj] {10.1086/115115}, \href
  {http://adsabs.harvard.edu/abs/1989AJ.....97.1752W} {97, 1752}

\bibitem[\protect\citeauthoryear{{Wo{\'z}niak} et~al.,}{{Wo{\'z}niak}
  et~al.}{2004}]{Wozniak:2004aa}
{Wo{\'z}niak} P.~R.,  et~al., 2004, \mn@doi [\aj] {10.1086/382719}, \href
  {http://adsabs.harvard.edu/abs/2004AJ....127.2436W} {127, 2436}

\bibitem[\protect\citeauthoryear{{Zharikov}, {Tovmassian}, {Aviles}, {Michel},
  {Gonzalez-Buitrago}  \& {Garc{\'{\i}}a-D{\'{\i}}az}}{{Zharikov}
  et~al.}{2013}]{Zharikov:2013aa}
{Zharikov} S.,  {Tovmassian} G.,  {Aviles} A.,  {Michel} R.,
  {Gonzalez-Buitrago} D.,   {Garc{\'{\i}}a-D{\'{\i}}az} M.~T.,  2013, \mn@doi
  [\aap] {10.1051/0004-6361/201220099}, \href
  {http://adsabs.harvard.edu/abs/2013A%26A...549A..77Z} {549, A77}

\makeatother
\end{thebibliography}

\end{document}